\documentclass[aps,prd,amsmath,amssymb,10pt,letterpaper,balancelastpage,showpacs,superscriptaddress,nofootinbib,floatfix,notitlepage,longbibliography,twocolumn
]{revtex4-2}
\usepackage[utf8]{inputenc}
\usepackage{bm}

\usepackage{comment}
\usepackage{slashed}
\usepackage{latexsym}
\usepackage{graphics}
\usepackage{bm}
\usepackage[dvipsnames]{xcolor}
\usepackage{amsmath}
\usepackage{amssymb}
\usepackage{mathrsfs}
\usepackage{graphicx}
\usepackage{slashed}
\usepackage{siunitx}
\usepackage{braket}
\usepackage{calligra}
\usepackage{caption}
\usepackage{subcaption}
\usepackage[referable]{threeparttablex}
\usepackage{threeparttable}

\usepackage{caption}
\usepackage{subcaption}

\newcommand{\fm}{f_{\rm MACHO}}
\newcommand{\Mm}{M_{\rm MACHO}}

\definecolor{plotgreen}{rgb}{0.0,0.66,0.0}
\definecolor{plotorange}{rgb}{1.0,0.5,0.0}
\definecolor{plotblue}{rgb}{0.0,0.0,1.0}
\definecolor{plotyellow}{rgb}{0.66,0.66,0.0}
\definecolor{plotbrown}{rgb}{0.6,0.4,0.2}
\definecolor{plotpurple}{rgb}{0.5,0.0,0.5}

\begin{document}
\title{Robust bounds on MACHOs from the faintest galaxies}

\date{\today}

\author{Peter W.~Graham} 
\email{pwgraham@stanford.edu}
\affiliation{Leinweber Institute for Theoretical Physics at Stanford, Department of Physics, Stanford University, Stanford, CA 94305, USA}
\affiliation{Kavli Institute for Particle Astrophysics \& Cosmology, Department of Physics, Stanford University, Stanford, CA 94305, USA}

\author{Harikrishnan Ramani} 
\email{hramani@udel.edu}
\affiliation{Department of Physics and Astronomy and the Bartol Research Institute, University of Delaware, Newark, DE 19716, USA}

\author{Maximilian Ruhdorfer} 
\email{m.ruhdorfer@stanford.edu}
\affiliation{Leinweber Institute for Theoretical Physics at Stanford, Department of Physics, Stanford University, Stanford, CA 94305, USA}

\begin{abstract}
We use the dynamical heating of stars in ultrafaint dwarf (UFD) galaxies to set limits on Massive Compact Halo Objects (MACHOs). In our analysis we study the robustness of the bounds under uncertainties in key UFD parameters, such as the half-light radius, stellar velocity dispersion, total halo mass and dark matter and stellar density profiles. We apply this framework to both well-established UFD candidates, as well as the recently discovered UFD candidate Ursa~Major~III/UNIONS~1. We find that multiple UFDs yield consistently strong limits in the mass range $10\, M_\odot \lesssim \Mm \lesssim 10^9\, M_\odot$, underscoring the robustness of a previous analysis solely based on Segue~I. We also demonstrate that Ursa~Major~III, if confirmed as an UFD, would improve the constraints significantly, providing the strongest constraints on MACHO dark matter in the mass range $1\, M_{\odot}\lesssim \Mm \lesssim 10^5\, M_\odot$.
\end{abstract}

\maketitle

\section{Introduction}
The identity of the dark matter (DM) particle remains unknown despite decades of theoretical and experimental investigation. While many particle candidates have been proposed, direct, indirect, and collider searches have placed increasingly stringent bounds on non-gravitational interactions between DM and the Standard Model. This motivates complementary approaches that exploit purely gravitational effects. The study of small scale structure is one such direction.   

Presently, structures are well studied down to wavenumbers of order $k \sim 10~\textrm{Mpc}^{-1}$, primarily through constraints derived from stellar tracers in gravitational halos down to the dwarf-galaxy scale. However, halos with masses $\ll 10^9~M_\odot$ typically fail to form stars, necessitating novel probes of DM substructure at these smaller scales. Many viable DM models make distinct predictions for structure formation on scales below those currently accessible to observations. Thus, finding new ways to study this small-scale structure would readily translate to testing the DM microphysics purely gravitationally. 

Ultrafaint dwarf (UFD) galaxies provide a perfect environment to test hitherto unconstrained length scales. Small scale structure, if present inside the UFD, would have caused dynamical heating of the central star cluster purely via gravitational scattering through its entire existence. This results in the expansion of its half-light radius. Thus, stringent limits can be placed on the UFD's substructure, if it results in the expansion of the half-light radius of the UFD to values that exceed the one observed today. UFDs are the ideal environment to observe this effect due to the following reasons.    Their dense DM halos and low velocity dispersions~\cite{Pace:2024sys} both enhance gravitational dynamical heating. The star population is some of the oldest observed, thus maximizing the time available for heat transfer. The half-light radii of the stellar populations of UFDs are also some of the smallest, thus making them highly sensitive to energy injection. These properties make UFDs powerful laboratories for constraining DM substructure, as demonstrated in prior work placing limits on Massive Compact Halo Objects (MACHOs)~\cite{Brandt:2016aco,Graham:2023unf,zhu2018primordial,Stegmann:2019wyz,zoutendijk2020muse,Wadekar:2022ymq,Koulen:2024emg,Lu:2020bmd,Koushiappas:2017chw}, ultralight dark matter (ULDM)~\cite{Marsh:2018zyw,Dalal:2022rmp,DuttaChowdhury:2023qxg,Yang:2024hvb,Teodori:2025rul,May:2025ppj,Lancaster:2019mde}, and diffuse substructure~\cite{Graham:2024hah}.  

MACHOs constitute macroscopic, dense objects whose mean densities exceed those of conventional substructure. Primordial black holes (PBHs) are the best-known example~\cite{Green:2020jor}, but other possibilities include axion stars (or more generally Bose stars)~\cite{Kolb:1993zz}, quark nuggets~\cite{Witten:1984rs}, and other compact states~\cite{coleman1985q,lee1992nontopological,Wise:2014jva,Grabowska:2018lnd} arising from DM microphysics. A scenario in which MACHOs account for all of DM is ruled out across the mass range relevant to this work ($M_{\rm MACHO} \gtrsim 1~M_\odot$)~\cite{EROS-2:2006ryy,Oguri:2017ock,Wilkinson:2001vv,10.1093/mnras/stac915,Ramirez:2022mys,moore1993upper,Carr:2020gox}. Nonetheless, MACHOs remain interesting as a possible sub-component, particularly since several of the above mentioned particle DM models naturally predict their production. Whereas PBHs are most strongly constrained by accretion and Hawking radiation, the broader class of MACHOs is constrained primarily via gravitational effects. 

In this work we extend the analysis of Ref.~\cite{Graham:2023unf} in two directions:  
\begin{enumerate}
    \item we quantify the effect of uncertainties in key UFD parameters such as total halo mass, halo density profile and stellar velocity disperson on heating-based bounds and
    \item we analyze many of the known UFD's including several newly discovered ones, with Ursa~Major~III/UNIONS~1 (if confirmed) providing the strongest constraints.   
\end{enumerate}

\section{Ultrafaint Dwarfs}\label{sec:UFDs}
In this work we analyze both well-established UFDs and the recently discovered candidate Ursa~Major~III/UNIONS~1~\cite{Smith:2023,Errani:2023sgd}. 
While there remains ongoing debate regarding whether Ursa Major III constitutes a genuine UFD rather than a self-gravitating stellar cluster~\cite{Devlin:2024}, if confirmed as an UFD it has an enormous potential to significantly improve stellar heating bounds due to its compact size and high DM density.

Table~\ref{tab:UFDList} summarizes all UFDs and their key properties considered in this work. For each UFD the 2D projected half-light radius $r_h$ and the stellar line-of-sight velocity dispersion $\sigma_{*,{\rm los}}$ are taken from~\cite{Pace:2024sys} and references therein. Following~\cite{Pace:2024sys,Stegmann:2019wyz} we take the age of all UFDs to be $T_{\rm UFD}=12$~Gyr and estimate the stellar mass by assuming a mass-to-light ratio of $2$, i.e. $M_{*}\simeq 2M_\odot/L_\odot \cdot L_V$, where $L_V$ is the observed V-band luminosity. We adopt the total halo masses $M_{\rm UFD}$ from~\cite{Stegmann:2019wyz} which were extracted from $N$-body simulations with PBH DM. Where data was available, we also checked that those values were within the error bars of fits of Milky Way UFDs to recasted Aquarius simulation results in~\cite{10.1093/mnras/sty2505}. The impact of total halo mass uncertainties on MACHO bounds will be quantified in Section~\ref{sec:bounds}.

\begin{table*}
\footnotesize
\vspace{30pt}
\caption{\label{tab:UFDList}\footnotesize List of UFDs considered in this work. Up-to-date values for the stellar line-of-sight velocity dispersion $\sigma_{*,{\rm los}}$ and the 2D projected stellar half-light radius $r_h$ are taken from~\cite{Pace:2024sys}. Estimates for the halo mass $M_{\rm UFD}$ are taken from $N$-body simulations performed in~\cite{Stegmann:2019wyz} and for the total stellar mass we assume a mass-to-light ratio of $2$, such that $M_{*}\simeq 2M_\odot/L_\odot \cdot L_V$, where $L_V$ is the observed V-band luminosity taken from~\cite{Pace:2024sys}. The values for Ursa~Major~III are taken from~\cite{Smith:2023,Errani:2023sgd}. The mean central density within the 3D half-light radius is computed according to Eq.~\eqref{eq:MeanDensity} which is based on the mass estimator in~\cite{Wolf:2009tu}. Note that for UFDs with only an upper bound on the velocity dispersion we took the velocity dispersions which were found for the simulations in~\cite{Stegmann:2019wyz} and listed in their Table~2. We indicate those values with $\spadesuit$ in the table. UFDs marked in boldface give the strongest MACHO heating bounds which are shown in Figure~\ref{fig:UFDBounds}.}
\begin{threeparttable}
\begingroup
\renewcommand*{\arraystretch}{1.3}
\begin{tabular}{lcccccc}
\hline
\hline
UFD Name & $\log \left(\frac{M_\text{UFD}}{M_\odot}\right)$ & $M_*$ & $r_{\rm h}$ & $\sigma_{\star,{\rm los}}$ & Mean Central Density\\
& ~ & ${M}_{\odot}$ & pc & km~s$^{-1}$ & $M_\odot \text{ pc}^{-3}$\\
\hline
{\bf Draco II} & $9.08^{+0.55}_{-0.47}$ & $3.6\times10^2$ & $19.0^{+4.5}_{-2.6}$ & $1.2 \,(<5.9$ at $95\%$ C.L.) $^\spadesuit$ &  $3.7\times 10^{-1}$\\ 

{\bf Segue I} & $9.00^{+0.56}_{-0.45}$ & $5.7\times10^2$ & $24.2\pm2.8$ & $3.7^{+1.4}_{-1.1}$ &  $2.2$\\

Tucana III & $9.08^{+0.42}_{-0.56}$ & $5.7\times 10^2$ & $34\pm8$ & $0.3~(<1.2$ at $90\%$ C.L.) $^\spadesuit$ &  $7.3 \times 10^{-3}$\\

{\bf Triangulum II} & $9.17^{+0.46}_{-0.56}$ & $5.7\times10^2$ & $17.4\pm4.3$ & $1.2~(<4.2$ at $90\%$ C.L.) $^\spadesuit$ & $4.5\times 10^{-1}$\\

Segue II & $8.94^{+0.52}_{-0.51}$ & $1.0\times10^3$ & $38.3\pm2.8$ & $0.6~(<2.6$ at $95\%$ C.L.) $^\spadesuit$ & $2.3\cdot 10^{-2}$\\

{\bf Carina III} & $8.94^{+0.63}_{-0.42}$ & $1.6\times10^3$ & $30\pm9$ & $5.6^{+4.3}_{-2.1}$  & $3.3$\\

{\bf Willman I} & $9.16^{+0.37}_{-0.66}$ & $1.8\times10^3$ & $27.7\pm2.4$ & $4.0\pm0.8$ & $2.0$\\

Bo\"{o}tes II & $9.02^{+0.46}_{-0.52}$ & $2.6\times10^3$ & $ 38.7\pm5.1$ & $2.9^{+1.60}_{-1.20}$ & $5.3\times 10^{-1}$\\

Grus I & $8.97^{+0.51}_{-0.52}$ & $7.7\times10^3$ & $151^{+21}_{-31}$ & $2.5^{+1.3}_{-0.8}$ & $2.6\times 10^{-2}$\\

{\bf Horologium I} & $9.02^{+0.50}_{-0.54}$ & $3.8\times10^3$ & $36.5\pm7.1$ & $4.9^{+2.8}_{-0.9}$ & $1.7$\\

Reticulum II & $9.21^{+0.29}_{-0.71}$ & $3.0\times10^3$ & $58\pm 4$ & $3.6^{+1.0}_{-0.7}$ & $3.6\times 10^{-1}$\\

%Tucana II & $8.95^{+0.47}_{-0.52}$ & $5.4\times10^3$ & $208$ & $3.8^{+1.1}_{-0.7}$ & $3.1\times 10^{-2}$\\

Pegasus III & $9.02^{+0.46}_{-0.55}$ & $8.0\times10^3$ & $118^{+31}_{-30}$ & $5.4^{+3.0}_{-2.5}$ & $2.0\times 10^{-1}$\\

Pisces II & $8.91^{+0.57}_{-0.44}$ & $8.8\times10^3$ & $69\pm8$ & $5.4^{+3.6}_{-2.4}$ & $5.7\times 10^{-1}$\\

Ursa Major II & $9.06^{+0.43}_{-0.58}$ & $1.0\times10^4$ & $129\pm 4$ & $6.7\pm1.4$ & $2.5\times 10^{-1}$\\

Aquarius II & $8.98^{+0.46}_{-0.54}$ & $9.5\times10^3$ & $159\pm24$ & $4.7^{+1.8}_{-1.2}$ & $8.2\times 10^{-2}$\\

Coma Berenices & $8.97^{+0.49}_{-0.53}$ & $8.9\times10^3$ & $72.1\pm3.8$ & $4.6\pm0.8$ & $3.8\times 10^{-1}$\\

Leo V & $8.90^{+0.58}_{-0.46}$ & $9.8\times10^3$ & $51.8\pm16.6$ & $2.3^{+3.2}_{-1.6}$  & $1.8\times 10^{-1}$\\

Carina II & $9.0^{+0.49}_{-0.54}$ & $1.2\times10^4$ & $91\pm8$ & $3.4^{+1.2}_{-0.8}$ & $1.3\times 10^{-1}$\\

Hydra II & $8.96^{+0.46}_{-0.53}$ & $1.9\times10^4$ & $ 59.2\pm10.9$ & $1.3~ (<4.5$ $90\%$ at C.L.) $^\spadesuit$ & $4.5\times 10^{-2}$\\

Hydrus I & $9.07^{+0.43}_{-0.57}$ & $1.3\times10^4$ & $53\pm4$ & $2.7\pm0.5$ & $2.4\times 10^{-1}$\\

Leo IV & $9.03^{+0.45}_{-0.59}$ & $1.6\times10^4$ & $114\pm12$ & $3.3\pm1.7$ & $7.9\times 10^{-2}$\\

Ursa Major I & $9.06^{+0.52}_{-0.59}$ & $1.9\times10^4$ & $234$ & $7.0\pm1.0$ & $8.4\times 10^{-2}$\\

Canes Venatici II & $8.98^{+0.50}_{-0.52}$ & $2.0\times10^4$ & $66.6\pm11.1$ & $4.6\pm 0.8$ & $4.5\times 10^{-1}$\\

Hercules & $9.03^{+0.42}_{-0.58}$ & $3.6\times10^4$ & $216\pm17$ & $5.1\pm0.9$ & $5.2\times 10^{-2}$\\

Bo\"{o}tes I & $9.07^{+0.41}_{-0.57}$ & $4.4\times10^4$ & $191\pm5$ & $4.6^{+0.8}_{-0.6}$  &  $5.4\times 10^{-2}$\\
\hline
{\bf Ursa Major III} & $9.00$ & $16$ & $3\pm 1$ &  $1.9^{+1.4}_{-1.1}$ (or $3.7^{+1.4}_{-1.0}$) & $38$ (or $1.4\times 10^{2}$)\\
\hline
\hline
\end{tabular}
\endgroup
\end{threeparttable}
\end{table*}
Due to the limited availability of kinematic stellar data, very little is known about the dark matter distribution in UFD halos. Neither the shape nor the total DM mass are currently experimentally accessible. Some UFDs seem to prefer a cuspy DM profile~\cite{Hayashi_2023}, but there will be no resolution between cuspy or cored profiles in the near future~\cite{Chang:2020rem}. To bracket this uncertainty we consider two extreme cases when modeling the DM profile: a cored Dehnen profile~\cite{10.1093/mnras/265.1.250} and a cuspy NFW profile~\cite{Navarro:1996gj}. The cored Dehnen profile is given by
\begin{equation}\label{eq:Dehnen}
\rho_{\rm DM}^{\rm Dehn} (r, R_{\rm UFD}) = \frac{3 M_{\rm UFD}}{4\pi R_{\rm UFD}^3}\left( 1 + \frac{r}{R_{\rm UFD}}\right)^{-4}\,,
\end{equation}
and is completely determined by the total halo mass $M_{\rm UFD}$ and the scale radius $R_{\rm UFD}$ below which the density is approximately constant. For the NFW profile
\begin{equation}\label{eq:NFW}
\rho_{\rm DM}^{\rm NFW} (r,R_{\rm UFD}) = \frac{M_{\rm UFD} f(c)}{4\pi R_{\rm UFD}^3}\left(\frac{r}{R_{\rm UFD}}\right)^{-1}\left(1+\frac{r}{R_{\rm UFD}}\right)^{-2}%\theta(c R_{\rm UFD} - r)
\,,
\end{equation}
where $f(c)^{-1} = \log (1+c) -c/(1+c)$, the scale radius determines where the inner slope of $\rho\sim r^{-1}$ transitions to $\rho\sim r^{-3}$. 
Following standard practice, we fix $c=c_{200}$, such that the mean enclosed density within $c_{200} R_{\rm UFD}$ equals $200$ times the present-day critical density.
Thus for the NFW profile the UFD mass is actually $M_{\rm UFD}=M_{200}$, i.e. the mass enclosed by a sphere with radius $c_{200} R_{\rm UFD}$.

We fix the UFD scale radius $R_{\rm UFD}$ in both cases using the luminosity-weighted squared velocity dispersion, which is given by~\cite{Wolf:2009tu}
\begin{equation}\label{eq:MassEstimator}
    \left\langle \sigma_{*,{\rm los}}^2\right\rangle \simeq \frac{G M_{1/2}}{3 R_{1/2,*}}\,,
\end{equation}
where $R_{1/2,*} \simeq 4/3 r_h$ is the 3D deprojected half-light radius and $M_{1/2}$ is the mass enclosed within the 3D half-light radius, which is to a good approximation completely determined by the DM profile and can be obtained analytically from Eqs.~\eqref{eq:Dehnen} and~\eqref{eq:NFW} as a function of $R_{\rm UFD}$ and $M_{\rm UFD}$. Combined with the half-light radius and velocity dispersion measurement, this determines the UFD scale radius $R_{\rm UFD}$.
\begin{figure}[t]
\centering
\includegraphics[width=\columnwidth]
{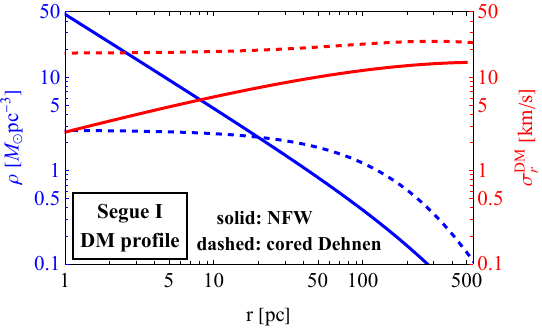}
\caption{\label{fig:segueProfile} DM density (blue) and radial velocity dispersion (red) for Segue~I, for an NFW (solid) and a cored Dehnen (dashed) profile. The profile parameters are fixed using the total halo mass $M_{\rm UFD} = 10^9\, M_\odot$ and the estimate of the stellar line-of-sight velocity dispersion in Eq.~\eqref{eq:MassEstimator}.  The velocity dispersions assume hydrostatic equilibrium and emphasize the different behavior of the velocity dispersion in NFW and cored Dehnen. While the velocity dispersion grows as $\sqrt{r}$ for an NFW profile, it is approximately constant within the core of a cored Dehnen profile.}
\end{figure}
As an example, we show the fit of an NFW (solid blue) and a cored Dehnen (dashed blue) profile to Segue~I in Figure~\ref{fig:segueProfile}. Despite their different shape, both profiles contain the same mass within the 3D half-light radius and predict the observed stellar velocity dispersion at the 3D half-light radius. However, the corresponding DM velocity dispersions in hydrostatic equilibrium have a fundamentally different behavior. While the velocity dispersion is approximately constant close to the center of the cored Dehnen profile, it decreases as $\sqrt{r}$ when approaching the center in the NFW profile. For $M_*\ll M_{\rm UFD}$ and $r\ll R_{\rm UFD}$ the velocity dispersions are approximately given by
\begin{equation}
\begin{split}
    \left. \sigma_{\rm DM}^2\right|_{\rm Dehn} &\simeq \frac{G\, M_{\rm UFD}}{30 R_{\rm UFD}}\,,\\
    \left. \sigma_{\rm DM}^2\right|_{\rm NFW} &\simeq g(c,r/R_{\rm UFD})\frac{G\, M_{\rm UFD}}{R_{\rm UFD}}\frac{r}{R_{\rm UFD}}\,,
\end{split}
\end{equation}
where $g(c,r/R_{\rm UFD})$ is a a mixed polynomial and logarithmic function of $c$ and $r/R_{\rm UFD}$ which is $\mathcal{O}(1)$ for typical values of $c$ and $r/R_{\rm UFD}$. Note that while in both cases the velocity dispersion seems to be directly sensitive to the total halo mass, even deep in the center of the UFD, this is only true for the cored Dehnen profile. There is an implicit dependence of $R_{\rm UFD}$ on $M_{\rm UFD}$ through the fitting procedure in Eq.~\eqref{eq:MassEstimator}. In the cored Dehnen profile $R_{\rm UFD} \sim M_{\rm UFD}^{1/3}$, what leaves a remnant dependence on the total halo mass in the velocity dispersion of $\left.\sigma_{\rm DM}^2\right|_{\rm Dehn}\sim M_{\rm UFD}^{2/3}$. However, in the NFW profile $R_{\rm UFD}\sim M_{\rm UFD}^{1/2}$, such that $\left. \sigma_{\rm DM}^2\right|_{\rm NFW}$ does not directly depend on the total halo mass.\footnote{There is still a small residual dependence through the concentration parameter $c_{\rm 200}$.} This intuitively makes sense since a decreasing radial velocity dispersion implies a small eccentricity of DM and MACHO orbits, such that the MACHOs with orbits close to the half-light radius, which predominantly contribute to the heating of the stars, are the ones which stay close to the galactic center and do not come from the edge of the galaxy.

We also use Eq.~\eqref{eq:MassEstimator} as an estimator for the mass within the 3D half-light radius to find the mean central density which we give in the last column of Table~\ref{tab:UFDList}
\begin{equation}\label{eq:MeanDensity}
    \bar{\rho} = \frac{M_{1/2}}{\tfrac{4\pi}{3}R_{1/2,*}^3} = \frac{9 \langle\sigma_{*,{\rm los}}^2\rangle }{4\pi G R_{1/2,*}^2}\,.
\end{equation}

For the stellar distribution, we primarily assume a Plummer profile~\cite{10.1093/mnras/71.5.460}
\begin{equation}
\rho_*^{\rm Plum} (r, R_{0,*}) = \frac{3 M_*}{4\pi R_{0,*}^3} \left( 1 + \left(\frac{r}{R_{0,*}}\right)^2 \right)^{-5/2}\,,
\end{equation}
where the stellar scale radius is given by $R_{0,*} = \tfrac{4}{3}\sqrt{2^{2/3}-1}\, r_h$ and $M_*$ is the total stellar mass given in Table~\ref{tab:UFDList}. In order to investigate the extent to which the bounds depend on the stellar profile, we also consider a second popular choice, an exponential stellar distribution
\begin{equation}
\rho_*^{\rm exp} (r, R_{0,*}) = \frac{M_*}{8\pi R_{0,*}^3} e^{-r/R_{0,*}}\,,
\end{equation}
where the scale-radius is given by $R_{0,*} \approx r_h / 2.02$.

\section{MACHO bounds from dynamical heating of stars}
In this section, we briefly review dynamical heating of the stellar population in UFDs due to gravitational scattering between stars and MACHOs. The discussion closely follows~\cite{Graham:2023unf,Graham:2024hah}, where the reader can find further details. Here we focus on modifications that arise due to a change in the stellar or DM profile.

We model UFDs as a coupled system of stars and DM, where the DM is composed of light particles of mass $m_{\rm DM}$ with a smooth density profile and MACHOs of mass $\Mm \gg m_{\rm DM}$ which make up a fraction $\fm$ of the total DM density.
There are two heating processes which are relevant for our discussion: i) MACHO migration: the MACHOs heat the smooth DM component, lose energy and migrate inwards towards the center of the galaxy, effectively enhancing the MACHO density that the stars encounter close to the half-light radius, and ii) stellar heating: the MACHOs heat the stars, which leads to an expansion of the stellar radius. While both processes are not independent, it is a good approximation to first consider MACHO migration and compute an enhanced effective MACHO density which we use to compute stellar heating~\cite{Graham:2023unf}.
%
%%%%%%%%%
\subsection{Migration of MACHOs}
%%%%%%%%%
%
We assume that the MACHOs and smooth DM initially have identical density profiles up to an overall normalization i.e.
\begin{equation}
    \rho_{\rm MACHO}(r,t=0)=f_{\rm MACHO} \rho_{\rm DM}(r,R_{\rm UFD})\,,
\end{equation}
with subsequent time-evolution entirely captured by a time dependent scale radius while keeping the overall shape the same i.e.
\begin{equation}
    \rho_{\rm MACHO}(r,t)=f_{\rm MACHO} \rho_{\rm DM}(r,R_{\rm MACHO}(t))\,,
\end{equation}
with $R_{\rm MACHO}(t=0)=R_{\rm UFD}$.
Since the MACHOs are in the smooth DM halo they will transfer kinetic energy to the smooth DM component which is the cause of the time dependence of the MACHO scale radius $R_{\rm MACHO}(t)$. In the limit $m_{\rm DM} \ll \Mm$, the heating rate of MACHOs per unit mass is given by~\cite{Graham:2023unf}\footnote{Note that the heating rate is negative since MACHOs lose energy to the smooth DM component.}
\begin{equation}\label{eq:MACHOHeating}
\begin{aligned}
H_{\mathrm{MACHO}} & =-2 \sqrt{2 \pi} G^2\left(1-f_{\mathrm{MACHO}}\right) \rho_{\mathrm{DM}} M_{\mathrm{MACHO}} \\
& \times \frac{\sigma_{\mathrm{MACHO}}^2}{\left(\sigma_{\mathrm{DM}}^2+\sigma_{\mathrm{MACHO}}^2\right)^{\frac{3}{2}}} \log \left(\frac{b_{90}^2+b_{\max }^2}{b_{90}^2+b_{\min }^2}\right)\,,
\end{aligned}
\end{equation}
where $G$ is the gravitational constant, $\sigma_{\rm MACHO}$ and $\sigma_{\rm DM}$ are the MACHO and smooth DM velocity dispersions, $b_{90}=G m_{\rm DM} M_{\rm MACHO}/(\mu v_{\rm rel}^2)$ is the impact parameter for a deflection of $90^\circ$ with $\mu = m_{\rm DM} M_{\rm MACHO}/(m_{\rm DM} + M_{\rm MACHO})$ and $b_{\rm min}$ and $b_{\rm max}$ are the minimal and maximal impact parameters of MACHO-smooth DM scattering for which we take $b_{\rm min} = 0$ and $b_{\rm max} = R_{\rm UFD}$. 

The change in the MACHO energy per unit mass is given by
\begin{equation}
\begin{split}
    \frac{d \mathcal{E}_{\rm MACHO} (R_{\rm MACHO})}{dt} = &\int_0^\infty dr\, 4\pi r^2 \rho_{\rm MACHO} (r,R_{\rm MACHO})\\ 
    &\,\times\, H_{\rm MACHO}(r,R_{\rm MACHO})\,,
\end{split}
\end{equation}
where the radius and $R_{\rm MACHO}$ dependence in $H_{\rm MACHO}$ originates from $\sigma_{\rm MACHO}(r,R_{\rm MACHO})$, $\sigma_{\rm DM}(r,R_{\rm MACHO})$ and $\rho_{\rm DM}(r, R_{\rm UFD})$. 
The direct dependence of the MACHO energy $\mathcal{E}_{\rm MACHO}(R_{\rm MACHO})$ on the scale radius $R_{\rm MACHO}$ can be determined by computing the total kinetic and potential energy of the system using the DM density profile. The expressions for a cored Dehnen and an NFW profile are given in Appendix~\ref{app:EnergyRadius}.

Using these results, we can compute the present-day MACHO scale radius. In order to quantify the enhancement of the MACHO number density due to migration at the location of the stars we define an effective enhancement factor $\eta$ which averages the relative change in the number density at the 3D half-light radius over the age of the UFD $T_{\rm UFD}$
\begin{equation}
\eta=\frac{1}{T_{\mathrm{UFD}}} \int_0^{T_{\mathrm{UFD}}} d t \frac{\rho_{\text {MACHO}}\left(R_{1/2, \star}, R_{\rm MACHO}(t)\right)}{\rho_{\text {MACHO }}(R_{1/2, \star}, R_{\rm MACHO}(0))}\,.
\end{equation}
In the following, we model the enhancement of the MACHO density through migration by rescaling $\fm\rightarrow \eta\, \fm$. Note that we assume that the smooth DM density profile does not change while the MACHOs migrate inwards. This is a good approximation as long as the MACHO energy density is much smaller than the smooth DM density such that the backreaction is negligible. For this reason we stop the evolution of $R_{\rm MACHO}$ at a radius $R_{\rm MACHO}^{\rm cut}$. In the cored Dehnen profile we choose this to be the radius at which the MACHO mass inside a sphere of this radius equals the smooth DM mass within the same sphere. This occurs at $R_{\rm MACHO}^{\rm cut} \approx \fm^{1/3}R_{\rm UFD}/2$. In the NFW profile we stop the evolution when the MACHO density at the center equals the smooth DM density $\rho_{\rm MACHO}(0)/\rho_{\rm DM}^{\rm smooth}(0)=1$, which occurs at $R_{\rm MACHO}^{\rm cut} = \fm^{1/2} R_{\rm UFD}$. This choice is conservative since it takes into account that due to the cusp in the center of the NFW profile, the central MACHO density could be significantly larger than the smooth DM density if only the mass and not the density in the central region is compared.
\begin{figure}[t]
\centering
\includegraphics[width=\columnwidth]
{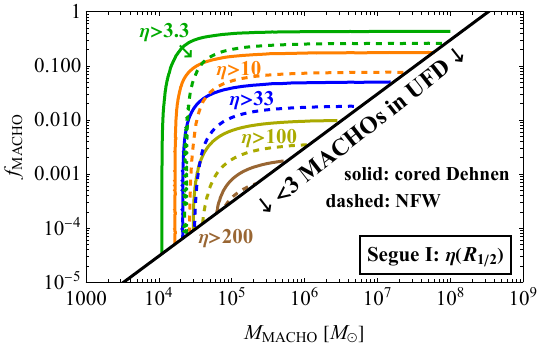}
\caption{\label{fig:Segue_migration} MACHO Density enhancement factor $\eta$ for Segue~I. The solid (dashed) lines show contours of constant $\eta$ assuming a cored Dehnen (NFW) DM profile. The parameter region below the black diagonal line corresponds to a scenario with on average less than $3$ MACHOs in the whole UFD and is therefore not experimentally accessible.}
\end{figure}
In Figure~\ref{fig:Segue_migration} we show contours of constant $\eta$ assuming a cored Dehnen (solid) and an NFW (dashed) profile. In both cases the enhancement factor increases with the MACHO mass $\Mm$. This is due to the increase of the heating rate in Eq.~\eqref{eq:MACHOHeating} with the MACHO mass. However, the increase occurs later in the NFW profile. 
The slightly counter-intuitive increase of the enhancement factor with decreasing $\fm$ is a consequence of our prescription to stop the evolution of the MACHO scale radius when the MACHO and smooth DM densities become comparable. This happens earlier for larger $\fm$, such that there is barely any migration of the MACHOs towards the center of the galaxy and consequently only a small density enhancement.

Before moving on, we comment on the interplay between migration and MACHO-MACHO scattering in cuspy profiles. As established earlier, MACHO scattering with the smooth DM causes migration, which results in the cusp getting sharper. However, it is also well known that MACHO-MACHO scattering, akin to self-interacting DM, causes the MACHO population to isothermalize, resulting in cusps becoming cores (see e.g.~\cite{Quinlan:1996bw,Zhu:2017plg,Boldrini:2019isx}). These are opposing effects and which one dominates depends on the MACHO fraction $f_{\rm MACHO}$. Assuming both the smooth DM and MACHOs follow the same density profile, MACHO-MACHO scattering is suppressed compared to migration by a factor $f_{\rm MACHO}$. A more refined estimate of the core formation and migration timescales reveals that up to an $\mathcal{O}(1)$ number the ratio of both timescales scales as
\begin{equation}
    \frac{T_{\rm core}}{T_{\rm mig}} \sim \frac{1-\fm}{\fm}\,,
\end{equation}
where the $T_{\rm mig} \sim (1-\fm )^{-1}$ scaling originates from $H_{\rm MACHO} \propto (1-\fm)$ in Eq.~\eqref{eq:MACHOHeating}.

Hence, for $\fm\lesssim 0.5$, migration dominates over core formation. However, both competing effects are always present. Especially when we stop migration, the MACHO and smooth DM densities are comparable close to the galactic center, such that locally $\fm \sim 0.5$ might be reached and core formation becomes an important effect during the late-time evolution of the UFD. Nevertheless, we find that there is no region in parameter space where this is relevant. Migration itself is only relevant for MACHO masses $\Mm \gtrsim 10^4\, M_\odot$, as can be seen in Figure~\ref{fig:Segue_migration}. As we will see in Section~\ref{sec:bounds}, in this heavy MACHO region the bound is saturated by sampling limitations, i.e. by the requirement that there are at least $3$ MACHOs in the UFD. Thus, an actually smaller enhancement factor due to core formation at the late stage of the UFD evolution would be compensated by the large heating rate, making the effect irrelevant for the limit-setting procedure. We also note that this intuitive understanding of the interplay between migration and core formation is consistent with the results in~\cite{Boldrini:2019isx}, where the authors performed simulations in which a fraction $\geq 0.01$ of all DM is made up of PBHs.
%
%%%%%%
\subsection{Stellar Heating}
%%%%%%
%
Similar to MACHO migration we assume that the stars maintain the shape of the stellar profile such that the effect of heating can be completely captured by a change in the stellar scale radius. This self-similar evolution has been confirmed in explicit simulations~\cite{Penarrubia:2025auj}.
Following~\cite{Graham:2023unf,Graham:2024hah} we split the heating rate into a direct heating and a tidal heating contribution, i.e.
\begin{equation}
    \frac{d\mathcal{E}_* (R_{0,*})}{dt} = \int_0^\infty dr\, 4\pi r^2 \rho_* (r, R_{0,*}) H_*(r,R_{0,*})\,,
\end{equation}
where $H_* = H_{*,{\rm direct}} + H_{*,{\rm tidal}}$. The dependence of the stellar energy on the stellar scale radius can be found from the total kinetic and potential energy of the system and depends on both the DM and stellar distribution. We collect the expression for all distributions used in this article in Appendix~\ref{app:EnergyRadius}. 

The direct heating contribution originates from the momentum transfer from MACHOs scattering on individual stars when they directly pass through the bulk of the stellar distribution. In complete analogy with the heating rate of smooth DM from MACHOs, the direct heating rate of stars from MACHOs is given by
\begin{equation}\label{eq:directHeating}
\begin{split}
    H_{\star, \text { direct }}=2 \sqrt{2 \pi} G^2 &\rho_{\mathrm{MACHO}} \frac{M_{\mathrm{MACHO}} \sigma_{\mathrm{MACHO}}^2-m_{\star} \sigma_{\star}^2}{\left(\sigma_{\star}^2+\sigma_{\mathrm{MACHO}}^2\right)^{\frac{3}{2}}}\\ 
    &\times \, \log \left(\frac{b_{90}^2+b_{\max }^2}{b_{90}^2+b_{\min }^2}\right)\,,
\end{split}
\end{equation}
where $m_{\star}$ and $\sigma_{\star}$ are the average star mass and the stellar velocity dispersion.
We choose $b_{\rm min}$ by requiring that at least 3 MACHO scattering events occur within the lifetime of the UFD at an impact factor of $b\geq b_{\rm min}$, i.e. we choose $b_{\rm min} = b_{\rm samp}$ where $b_{\rm samp}$ satisfies
\begin{equation}\label{eq:bsamp}
\pi b_{\rm samp}^2 \frac{\fm \rho_{\rm DM}}{\Mm} v_{\rm MACHO} T_{\rm UFD}=3\,.
\end{equation}
Note that while $b_{\rm samp}$ is approximately constant in cored profiles, such as the cored Dehnen profile used in~\cite{Graham:2023unf}, it is position dependent in general, i.e. $b_{\rm samp} = b_{\rm samp}(r)$. The maximal impact parameter is chosen such that the passage time of the MACHO is smaller than the orbital period of the star around the UFD, since a long scattering event dilutes the amount of energy that is transferred to the star. At a distance $r$ from the center of the galaxy this condition implies
\begin{equation}
\frac{b_{\rm max}}{\sigma_{\rm MACHO}} \ll \frac{r}{\sigma_*}\,.
\end{equation}
Thus for stars at a radius $r$ from the galactic center we take
\begin{equation}\label{eq:bmax}
    b_{\rm max}(r) = \max \left(1, \frac{\sigma_{\rm MACHO}}{\sigma_*} \right)\cdot r\,.
\end{equation}
For a cored DM profile, such as the cored Dehnen in~\cite{Graham:2023unf}, with the majority of stars located close to the half-light radius this effectively simplifies to $b_{\rm max} = R_{1/2,\star} \sigma_{\rm MACHO} / \sigma_* $ when the integral over the stellar density is taken into account.
\begin{figure}[t]
\centering
\includegraphics[width=\columnwidth]
{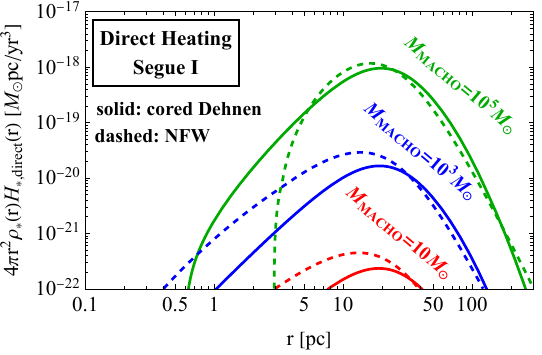}
\caption{\label{fig:DH_comp} Direct heating rate multiplied by the stellar distribution function for Segue~I assuming a cored Dehnen (solid) and an NFW (dashed) DM profile. The MACHO fraction is fixed to $\fm = 10^{-2}$ and we use a Plummer stellar profile with $M_* = 5.7\cdot 10^2\, M_\odot$ and $R_{0,*}=24.7$~pc in both cases.}
\end{figure}

The DM profile has an important impact on the direct heating rate. For cored DM profiles, such as the cored Dehnen profile that we consider here, the heating rate is approximately constant in the vicinity of the stars. In cuspy DM profiles, such as NFW, the direct heating rate has a non-trivial position dependence. A comparison of the direct heating rate multiplied by a Plummer stellar distribution for Segue~I, assuming a cored Dehnen or an NFW DM profile is shown in Figure~\ref{fig:DH_comp}. It is clearly visible that for larger MACHO masses $\Mm \gtrsim 10^4\, M_\odot$ the heating rate starts to get larger and eventually dominates in cored profiles, whereas for lighter masses heating in cuspy profiles dominates especially close to the center. This is due to a combination of two effects. Firstly, due to the cuspy nature of the NFW profile, the DM and therefore also the MACHO density grows towards the center of the galaxy, which leads to an enhancement in the direct heating rate, as can be seen in Eq.~\eqref{eq:directHeating}. This enhancement of the heating rate towards small $r$ in the NFW profile is clearly visible for light MACHO masses in Figure~\ref{fig:DH_comp}. 
However, this enhancement is countered by a second effect which is due to the logarithm in Eq.~\eqref{eq:directHeating}. When $b_{\rm max} = b_{\rm min}$ the logarithm and consequently the heating rate vanishes since there is no viable range of impact parameters. Towards the center of the UFD the MACHO velocity dispersion falls off as $\sigma_{\rm MACHO}\sim \sqrt{r}$ in the NFW profile, such that $b_{\rm samp}^2 \propto \Mm r^{-1/2}$ and $b_{\rm max}^2\propto r^2$, as can be seen from Eqs.~\eqref{eq:bsamp} and~\eqref{eq:bmax}. Thus, at some distance $r_{\rm min}$, $b_{\rm max}$ will equal $b_{\rm samp}$, making the heating rate zero. Due to the scaling of $b_{\rm samp}$ with $\Mm$ this happens earlier for larger MACHO masses and eventually so early that there is no enhancement from the larger MACHO density in the center of the galaxy. This is what happens for the green curve with $\Mm = 10^5 M_\odot$ in Figure~\ref{fig:DH_comp}, which barely reaches the heating rate that can be observed in a cored profile for the same MACHO mass. Also note that thanks to this screening the heating rate is insensitive to the unphysical singularity at the center of the NFW profile. The bulk of the heating rate has support close to the half-light radius. From this discussion we can conclude that stellar heating in UFDs with cuspy DM profiles is more sensitive to lighter MACHOs, whereas UFDs with cored profiles are an ideal probe for heavier MACHOs.

For impact parameters $b \gtrsim b_{\rm max}$ the MACHOs do not scatter off single stars anymore but they scatter coherently off the star cluster as a whole. This mostly leads to an acceleration of the center of mass of all stars and does not have a heating effect. The heating of the star cluster is caused by tidal effects. Similarly to~\cite{Graham:2023unf} we model tidal heating in the straight-line approximation for MACHOs that pass by the star cloud and heat it through tidal effects. For large impact parameters $b\gg R_{1/2.*}$ we can use the distant tide approximation~\cite{Gnedin:1997vp,vandenBosch:2017ynq} in which the energy transferred to stars per unit stellar mass from one pass of a MACHO is given by~\cite{vandenBosch:2017ynq}
\begin{equation}\label{eq:distantTide}
\begin{split}
\frac{\Delta E_{\rm dt}}{M_*} = \frac{4 G^2 M_{\rm MACHO}^2}{b^4 v_{\rm rel}^2}\frac{\langle r^2\rangle_*}{3}\,,
\end{split}
\end{equation}
where $\langle r^2\rangle_*$ is $r^2$ weighted by the stellar distribution
\begin{equation}
    \langle r^2\rangle_* = \frac{1}{M_*} \int_0^{r_{\rm max}} 4\pi r^4 \rho_* (r) dr\,.
\end{equation}
Motivated by~\cite{Graham:2023unf} we take $r_{\rm max} = 2.5 R_{1/2,*}$. Other choices change the result only by an $\mathcal{O}(1)$ number. Multiplying this by the flux of MACHOs and integrating over impact parameters, we obtain the tidal heating rate 
\begin{equation}
\begin{split}
H_{*,{\rm tidal}}(r) = \int_{b_{\rm min}^{\rm tidal}}^{b_{\rm max}^{\rm tidal}} db&\, 2\pi b\, A_{\rm corr}(b) \frac{\Delta E_{\rm tidal}(b)}{M_*} \\
&\times\, \frac{\rho_{\rm MACHO}}{M_{\rm MACHO}} v_{\rm MACHO}\,,
\end{split}
\end{equation}
where we left the $r$-dependence of all quantities implicit and identified $\Delta E_{\rm tidal}$ with $\Delta E_{\rm dt}$ for now. We also added an adiabatic correction factor $A_{\rm corr}(b)$ which is given by~\cite{vandenBosch:2017ynq} 
\begin{equation}
A_{\text {corr }} (r,b)=\left(1+\left(\frac{\sigma_{\star}\left(r\right)}{r} \frac{b}{\sigma_{\text {MACHO }}(r)}\right)^2\right)^{-\frac{3}{2}}\,.
\end{equation}
The adiabatic correction factor takes into account that the heating rate is lowered for scattering times that are larger than the orbital frequency of the stars. As the maximal impact parameter we choose the scale radius of the UFD, i.e. $b_{\rm max}^{\rm tidal} = R_{\rm UFD}$. However, the choice for the minimal impact parameter is not as straightforward. If we choose $b_{\rm min}^{\rm tidal} = b_{\rm max}$ we will in general include regions where the distant tide approximation is not appropriate since for $r\ll R_{1/2,*}$ typically $b_{\rm max}\ll R_{1/2,*}$, in which case Eq.~\eqref{eq:distantTide} overestimates the energy transfer as it grows as $b^{-4}$ for small impact parameters. Reference~\cite{vandenBosch:2017ynq} suggests to cut off the distant tide energy transfer at an impact parameter where it agrees with the energy transfer for a head-on collision with $b=0$. The energy transfer for the head-on collision is given by~\cite{vandenBosch:2017ynq}
\begin{equation}
\Delta E_{\rm ho} = \frac{4 G^2 M_{\rm MACHO}^2 \pi}{v_{\rm rel}^2} \int_0^{r_s} I_0^2(R) \Sigma_s (R) \dfrac{dR}{R}\,,
\end{equation}
where $\Sigma_s (R) = \int_{-\infty}^{+\infty}dz \rho_* (\sqrt{R^2 + z^2})$ is the projected surface density profile of the star cluster and the definition of $I_0$ can be found in~\cite{vandenBosch:2017ynq,gnedin1999tidal}. With this we define the tidal heating energy transfer as
\begin{equation}
\frac{\Delta E_{\rm tidal}}{M_*} = \frac{4 G^2 M_{\rm MACHO}^2 \langle r^2\rangle_* }{3 v_{\rm rel}^2} \times\begin{cases}
    b^{-4}\,, \quad b>b_0\\
    b_0^{-4}\,,\quad b\leq b_0
\end{cases}\,,
\end{equation}
where $b_0$ is defined as the impact parameter where $\Delta E_{\rm dt} (b_0) = \Delta E_{\rm ho}$. After this modification we can choose $b_{\rm min}^{\rm tidal} = \text{max}(b_{\rm max}, b_{\rm samp})$. We take the maximum of $b_{\rm max}$ and $b_{\rm samp}$ since $b_{\rm samp}$ can be larger than $b_{\rm max}$ especially for large MACHO masses. Note that for all scenarios that we consider, tidal heating is subleading to direct heating after MACHO migration has been included.
%
%%%%
\subsection{Stellar Heating Bounds}\label{sec:bounds}
%%%%
%
Now we apply this formalism to systematically derive bounds on MACHO DM. In~\cite{Graham:2023unf} Segue-I was used to set bounds assuming a cored Dehnen DM and Plummer stellar profile. Here we first revisit Segue-I and study how the bound is affected by the choice of DM and stellar profiles and how uncertainties on the astrophysical parameters modify the bound. Finally, we present the bounds from the list of UFDs in Table~\ref{tab:UFDList}.

The central idea behind heating bounds is that the energy injected by gravitational scattering between MACHOs and stars leads to an expansion of the stellar half-light radius during the lifetime of the UFD. The observed half-light radius today therefore sets a limit on the allowed amount of heating or equivalently the maximally allowed abundance of MACHO DM.
\begin{figure}[t]
\centering
\includegraphics[width=\columnwidth]
{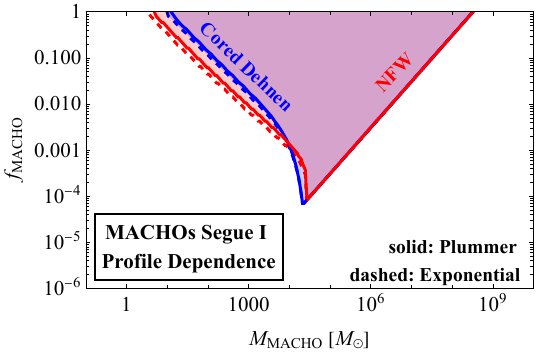}
\caption{\label{fig:SegueStellar} 
Profile dependence of the limits on the fraction of MACHO DM $f_{\rm MACHO}$ as a function of the MACHO mass $M_{\rm MACHO}$ from dynamical heating of stars in Segue~I. In red (blue) we show the limits assuming an NFW (cored Dehnnen) profile for DM.  The solid and dashed curves show results for a Plummer and exponential profile for the stars, respectively.
}
\end{figure}

Thus, in order to set a bound we have to determine the evolution of the stellar scale radius due to heating from MACHOs. In order to do so we use the results of the previous section to find the full stellar heating rate $H_* = H_{*, {\rm direct}} + H_{*,{\rm tidal}}$, which is required to obtain the evolution of the stellar scale radius
\begin{equation}\label{eq:RstarEvolution}
    \frac{d R_{0,*}}{dt} = \left[ \frac{d\mathcal{E}_*}{d R_{0,*}}\right]^{-1} \int_0^\infty dr\, 4\pi r^2 \rho_*(r,R_{0,*}) H_* (r, R_{0,*})\,,
\end{equation}
where $d\mathcal{E}_* / dR_{0,*}$ is determined from the potential and kinetic energy of the stellar system coupled to the DM halo. More details and explicit expressions for all profiles used in this work are given in Appendix~\ref{app:EnergyRadius}. We solve this differential equation by requiring that the stellar radius today matches the observed half-light radius, i.e. $R_{0,*} (T_{\rm UFD}) = R_{0,*}^{\rm today}$. For Segue-I, which has a 2D projected half-light radius of $r_h = 24.2$~pc, a line-of-sight stellar velocity dispersion of $\sqrt{\langle \sigma_{*,{\rm los}}^2\rangle} = 3.7$~km/s, a total halo mass of $M_{\rm UFD} = 10^9 M_\odot$ and a total stellar mass of $M_* = 5.7\cdot 10^2 M_\odot$,\footnote{Note that in~\cite{Graham:2023unf,Graham:2024hah} a total stellar mass of $M_*=10^3 M_\odot$ was taken. Here we take $M_* = 5.7\cdot 10^2 M_\odot$ since it corresponds to a mass-to-light ratio of $2$ which we assume for all UFDs. This has a negligible effect on the bound.} the stellar scale radius today is $\left. R_{0,*}^{\rm today}\right|_{\rm Exp} = 12.0$~pc assuming an exponential stellar profile and $\left. R_{0,*}^{\rm today}\right|_{\rm Plum} = 24.7$~pc for a Plummer profile. 
\begin{figure*}[t]
\centering
\includegraphics[width=0.49\textwidth]
{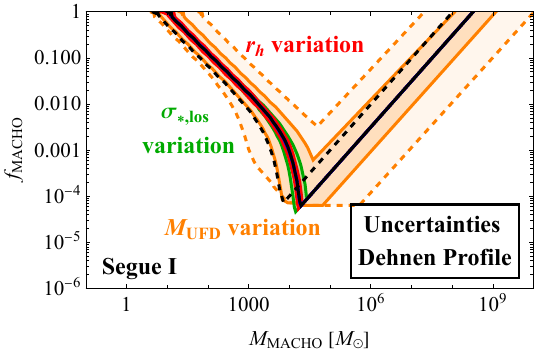}
\hfill
\includegraphics[width=0.49\textwidth]
{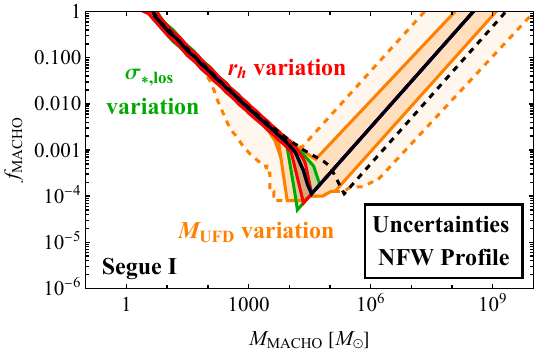}
\caption{\label{fig:UFDBoundsUncertainty} Dependence on the uncertainty of the astrophysical input parameters of the limits on the fraction of MACHO DM $f_{\rm MACHO}$ as a function of the MACHO mass $M_{\rm MACHO}$ from dynamical heating of stars in Segue-I. The figure shows the variation of the bound within the uncertainties of the stellar half-light radius (red), the line-of-sight velocity dispersion (green) and the total halo mass (orange). The solid black line assumes the central values of all parameters from \cite{Stegmann:2019wyz}. The solid orange lines correspond to the total halo mass uncertainties provided in~\cite{Stegmann:2019wyz}.  The dashed black and orange lines are the central value and uncertainties of the total halo mass from~\cite{10.1093/mnras/sty2505}. See the text for details. The {\bf left} panel shows the variation of the bound in a cored Dehnen profile, whereas the {\bf right} panel assumes an NFW DM profile. The bounds in both plots are for a Plummer stellar profile.}
\end{figure*}

For each $\fm$ and $\Mm$ we take the effect of MACHO migration into account by substituting $\fm \rightarrow \eta\, \fm$ in the heating rate and compute the initial stellar scale-radius $R_{0,*}(t=0)$ that would give the observed $R_{0,*}^{\rm today}$. A given parameter point is excluded if the initial scale-radius falls below a threshold value $R_{0,*}^i$, corresponding to an initial stellar core density of $\rho_*^i = \rho_*(R_{0,*}^i) \approx 5.27 M_\odot \text{pc}^{-3}$ which is comparable to some of the densest stellar clusters that we have observed. For Segue-I this is $\left.R_{0,*}^i\right|_{\rm Plum} = 1.66$~pc, assuming a Plummer profile and $\left.R_{0,*}^i\right|_{\rm Exp} = 1.17$~pc for an exponential profile. Note that as already pointed out in~\cite{Graham:2023unf}, the bounds depend only very weakly on $R_{0,*}^i$. We checked that increasing or decreasing $R_{0,*}^i$ by a factor of $3$ has a negligible effect on the bound.

The resulting bounds for all combinations of the DM and stellar profile are shown in Figure~\ref{fig:SegueStellar}. 
The Figure confirms what was already visible for the direct heating rate in Figure~\ref{fig:DH_comp}: there is more stellar heating for low MACHO masses $\Mm \lesssim 10^4 M_\odot$ in a cuspy profile, such as NFW (red), than in a cored profile, which results in stronger bounds. For heavier MACHOs the heating rate is reduced in cuspy DM profiles despite the denser core due to the increase of the lower bound on the impact parameters $b_{\rm min} = b_{\rm samp}$ with the MACHO mass. In this region of parameter space, cored DM halos such as those described by a cored Dehnen profile (blue) give stronger bounds. For both profiles the right edge of the bound is determined by requiring that there are at least $3$ MACHOs in the UFD, i.e. $\fm M_{\rm UFD} > 3\Mm$. The left edge is determined by the direct heating component, which dominates over tidal heating at low masses. In the cored profile the densities and velocity dispersions are approximately constant within the half-light radius, such that $H_{*, {\rm direct}} \propto \fm \Mm$, which causes the left edge of the bound to behave as $\fm \propto \Mm^{-1}$ until MACHO migration boosts the heating rate. This occurs roughly for $\Mm \gtrsim 10^3 M_\odot$, leading to a sudden change of the slope of the bound at $\Mm\simeq 10^3 M_\odot$. The earlier onset of migration at lower masses in the cored Dehnen profile is the cause for the stronger bounds at masses $\Mm \gtrsim 10^3 M_\odot$. Thus, assuming a cored Dehnen profile does not in general give conservative bounds. The bounds are only conservative for low masses.
Overall though, the bounds are not greatly affected by the choice of DM profile as seen in Figure~\ref{fig:SegueStellar}.
The stellar profile has only a minor effect on the bound. Nonetheless, the Plummer profile (solid) consistently leads to slightly weaker bounds than the exponential profile (dashed), irrespective of the DM density profile. The difference comes mainly from the dependence of the energy of the stellar system on the stellar scale radius, i.e. $d\mathcal{E}_*/d R_{0,*}$ in Eq.~\eqref{eq:RstarEvolution} (see Appendix~\ref{app:EnergyRadius}). 
While the bound is almost insensitive to the stellar profile in the cored Dehnen halo, in the NFW DM profile we find that the bound improves by a factor of $1.35$ for the exponential stellar profile over the Plummer profile. 
In the following we will make the conservative choice to always assume a Plummer stellar profile when deriving bounds. Before moving on, let us stress again that the overall profile dependence of the bound is small, making the bounds robust, regardless of what DM density profile is realized in UFDs.

For all bounds so far, we always took the central value for the astrophysical input parameters, as listed in Table~\ref{tab:UFDList}. However, the half-light radius, the line-of-sight velocity dispersion and above all the UFD mass, which can only be inferred from fits to simulations, come with a considerable uncertainty. In Figure~\ref{fig:UFDBoundsUncertainty} we show the bounds which are obtained by individually varying these input parameters within their experimental uncertainties as given in Table~\ref{tab:UFDList}. % and assuming a Plummer stellar profile. 
For the total stellar mass we vary the mass-to-light ratio from $1$ to $3$. We find that the dependence of the bound on the total stellar mass is too weak to be visible in the plot. The variation of the half-light radius (red) and the velocity dispersion (green) has only a mild effect on the bound, which gets more pronounced for larger MACHO masses.  However, the variation of the total halo mass (orange) has a significant effect on the bound, especially the high mass part (right side). This is mainly due to the large relative uncertainties of more than $100\%$ associated with it. The orange-shaded region bounded by the solid orange curves depicts the variation of the bound within the total halo mass uncertainty given by~\cite{Stegmann:2019wyz} $M_{\rm UFD} = 10^{9.0^{+0.56}_{-0.45}}M_\odot$. 
However, the mass estimate and uncertainty seem strongly correlated with the prior distribution for the total halo mass (a log normal distribution $\log_{10} (M_{\rm UFD}/M_\odot) \sim \mathcal{N}(9.0;0.5)$) which was used for their simulations. 
Thus, the quoted uncertainties may underestimate the real uncertainties. For this reason we also consider the mass estimate and its uncertainties from~\cite{10.1093/mnras/sty2505}. Ref.~\cite{10.1093/mnras/sty2505} separately fitted cored and cuspy DM profiles to observed data from Segue~I, using recasts of the Aquarius simulation to break the degeneracy of the total DM halo mass and the scale radius of the halo. They find $M_{\rm UFD}^{\rm core} = 10^{8.5^{+2.0}_{-0.7}} M_\odot$ for the cored profile and $M_{\rm UFD}^{\rm cusp} = 10^{9.8^{+0.6}_{-2.5}} M_\odot$ for the cuspy profile. They also point out that the cored profile provides a poor fit for Segue-I. We also note that a clear preference of Segue-I for a cuspy DM profile was also found in~\cite{Hayashi_2023}. We irrespectively use these values to estimate the effect of the uncertainty on the bound. The bound obtained using the central value of the DM mass from~\cite{10.1093/mnras/sty2505} is shown in dashed black and the variation within the uncertainty in dashed orange lines and the light orange shaded region. The variation of the total halo mass has only a mild effect on the low-mass region, where stellar heating is dominated by direct heating. Interestingly, the bound in the NFW profile seems to be completely insensitive to the total halo mass. This is a consequence of the observation in Section~\ref{sec:UFDs} that the DM velocity dispersion does not depend on $M_{\rm UFD}$, which makes the direct heating rate in Eq.~\eqref{eq:directHeating} insensitive to the total halo mass. In addition, the $M_{\rm UFD}$ dependence in the stellar energy per unit mass in Eq.~\eqref{eq:EpsStarNFWPlum} drops out due to the implicit halo mass dependence in the fit of $R_{\rm UFD}$, making the full half-light radius evolution equation and consequently the bound $M_{\rm UFD}$ independent of $M_{\rm UFD}$. Intuitively, this can be understood by the radial DM velocity dispersion which decreases towards the galactic center in the NFW profile. This implies that the orbits of MACHOs close to the stars have a small eccentricity and stay close to the center of the UFD. These are the MACHOs which are responsible for the direct heating of the stars. Thus, the stars are only sensitive to the central MACHO population. The biggest effect of varying the total halo mass within these uncertainties is a shift of the right edge of the bound. The right edge is determined by the requirement that the UFD contains at least $3$ MACHOs, i.e. by the requirement that $\fm M_{\rm UFD} > 3\Mm$. More massive UFDs can therefore probe heavier MACHOs. This implies that while the bounds are reasonably robust for lower MACHO masses, the bound at higher masses is highly dependent on the total halo mass. However, note that the central values for the total halo mass from both references are much closer than the uncertainties in~\cite{10.1093/mnras/sty2505}. 

\begin{figure*}[t]
\centering
\includegraphics[width=0.49\textwidth]
{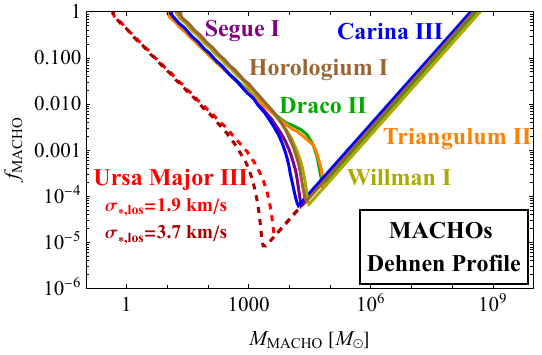}
\hfill
\includegraphics[width=0.49\textwidth]
{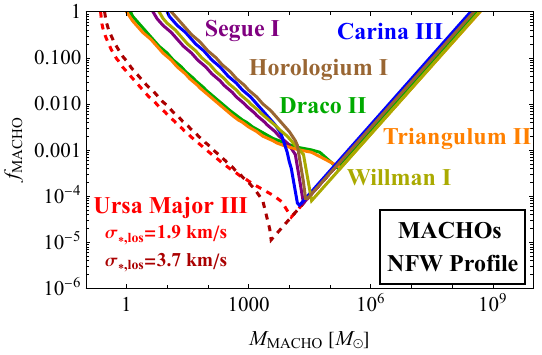}
\caption{\label{fig:UFDBounds} Limits on the fraction of MACHO DM $f_{\rm MACHO}$ as a function of the MACHO mass $M_{\rm MACHO}$ from dynamical heating of stars. Shown in solid lines are the limits for the UFDs in Table~\ref{tab:UFDList} which give the strongest bounds. The red dashed and dark red dashed lines are projections for the bounds from Ursa~Major~III, assuming a velocity dispersion of $\sigma_{*,{\rm los}}=1.9$~km/s and $\sigma_{*,{\rm los}}=3.7$~km/s, respectively. The {\bf left} panel shows the bounds assuming a Dehnen and the {\bf right} panel assuming an NFW DM profile. Both panels assume a Plummer sphere for the stellar density.}
\end{figure*}

Since the bounds depend so strongly on the  total halo mass which is not directly observable, one should not rely on the bound from just a single UFD. However, if a handful of UFDs with similar central values of the inferred total halo mass yield similar bounds, the bound becomes much more robust since, on the one hand, it is highly likely that one UFD will have a true total halo mass that lies close to the central value as the total number of observed UFDs increases, and this is enough to set a bound.
On the other hand, since the right edge of the bound is set by sampling limitations, a large number of UFDs provide a larger combined total mass that overcomes the sampling limitations even if all UFDs have total halo masses close to the lower end of their uncertainty interval. However, note that no matter how many UFDs are observed, a combination of all observations cannot extend the bound to MACHO masses that are larger than the total halo mass of the individual UFDs. 

For this reason we evaluate the bounds from individual UFDs in Table~\ref{tab:UFDList}. The strongest bounds assuming a Dehnen or NFW profile, both with a Plummer stellar profile, are shown as solid lines in Figure~\ref{fig:UFDBounds} and highlighted in bold in Table~\ref{tab:UFDList}. All bounds that we show are comparable to the one from Segue~I, establishing that Segue~I is not an outlier and making its bound considerably more robust. The bounds from all remaining UFDs in Table~\ref{tab:UFDList} are shown in Figure~\ref{fig:AllUFD} in Appendix~\ref{app:AllUFDBound}.
%
%%%%
\subsection{Stellar Heating Bounds from Ursa~Major~III/UNIONS~1}
%%%%
%
As can be seen from Table~\ref{tab:UFDList} and Figure~\ref{fig:UFDBounds}, UFDs with a larger mean central density generically give stronger bounds. 
This is a strong indication that the recently observed UFD candidate Ursa~Major~III/UNIONS~1~\cite{Smith:2023} with a 2D projected half-light radius of $r_h = (3\pm 1)$~pc, a stellar mass of $M_* = 16^{+6}_{-5} M_\odot$ and a stellar line-of-sight velocity dispersion of $\sigma_{*,{\rm los}} = 3.7^{+1.4}_{-1.0}$~km/s or $\sigma_{*,{\rm los}}= 1.9^{+1.4}_{-1.1}$~km/s, might be be an ideal candidate for setting strong bounds on MACHO DM. Despite the large uncertainty on its velocity dispersion, where the first value includes all $11$ observed stars and the second value omits one star that appears to be an outlier in the observed velocity distribution, its inferred mean central density is at least an order of magnitude larger than that of any other UFD that we have considered so far. While its final confirmation as an UFD is still pending, there is already a range of circumstantial evidence supporting that Ursa~Major~III is indeed a DM dominated UFD and not a self-gravitating star cluster. Its old stellar population suggests an age of $T_{\rm UFD} > 10$~Gyr~\cite{Smith:2023} which would be unusually high for a stellar cluster in the tidal field of the Milky Way.
$N$-body simulations suggest that such a long survival of Ursa~Major~III %in the tidal field of the Milky Way 
can be explained if it is embedded in a massive cuspy DM halo~\cite{Errani:2023sgd}. 

A recent paper~\cite{Devlin:2024} suggests that Ursa~Major~III could be the remnant of an initially significantly more massive stellar cluster that was stripped by the tidal field of the Milky Way. Being such a remnant, they argue that the cluster could survive for $1-2$~Gyr, which would imply that we observe the cluster during a short window at the end of its life. However, as was pointed out in~\cite{May:2025ppj}, losing of the order of $6000$ stars during its lifetime would most likely leave an observable stellar stream that traces the motion of Ursa~Major~III through the Milky Way. The non-observation of such a stellar stream challenges the proposal that Ursa~Major~III originates from a more massive stellar cluster. Lastly, simulations~\cite{manwadkar2022forward} indicate that faint UFDs that match the properties of Ursa Major III should occur in close proximity to the Milky Way, providing strong theoretical motivation for the presence of extremely low-luminosity UFDs.

Thus, Ursa~Major~III being an UFD currently seems to be the more likely option. Following~\cite{Errani:2023sgd} we therefore model Ursa~Major~III as being composed of an NFW DM halo with mass $M_{\rm UFD}=10^9 M_\odot$. In the right panel of Figure~\ref{fig:UFDBounds} we show in dashed red lines the bound that the observation of Ursa~Major~III sets on MACHO DM. We show the bound for both options of the velocity dispersion in red and dark red, respectively. In the left panel, we also show the bounds assuming a cored Dehnen profile. In both cases the stars are distributed in a Plummer sphere. Except for the region $\Mm \lesssim 10^3 M_\odot$, the larger stellar line-of-sight velocity dispersion generically gives slightly stronger bounds since a larger stellar velocity dispersion implies a more compact UFD or equivalently a denser DM core. We therefore take $\sigma_{*,{\rm los}} = 1.9$~km/s as a conservative choice in the following and show bounds for the NFW profile, motivated by the $N$-body simulations in~\cite{Errani:2023sgd}.
\begin{figure}[t]
\centering
\includegraphics[width=\columnwidth]
{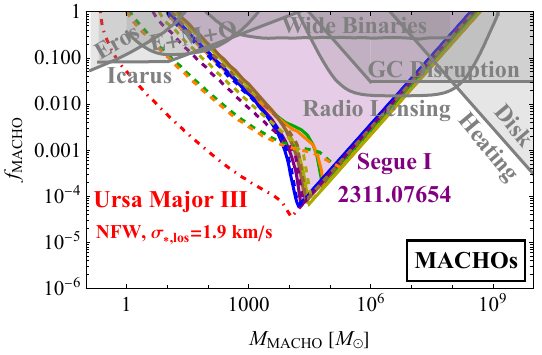}
\caption{\label{fig:FinalBounds} Bounds on the fraction of MACHO DM. The purple shaded region is the limit from stellar heating in Segue~I from~\cite{Graham:2023unf}. The red dot-dashed line shows the bound that can be obtained from Ursa~Major~III if it is confirmed as an UFD. The bound assumes an NFW DM profile, inspired by~\cite{Errani:2023sgd}, and takes the more conservative value for the line-of-sight stellar velocity dispersion of $\sigma_{*,los} = 1.9$~km/s. The colored solid and dashed lines show further stellar heating bounds from the UFDs {\color{plotpurple} Segue~I}, {\color{plotgreen} Draco~II}, {\color{plotorange} Triangulum~II}, {\color{plotblue} Carina~III}, {\color{plotyellow} Willman~I}, {\color{plotbrown} Horologium~I}, assuming a Dehnen and NFW DM profile, respectively. All stellar heating bounds are derived for stars distributed according to a Plummer sphere. Further limits from microlensing (EROS)~\cite{EROS-2:2006ryy}, caustic crossings of Icarus~\cite{Oguri:2017ock}, multiple imaging from radio sources~\cite{Wilkinson:2001vv,10.1093/mnras/stac915}, non-disruption of wide binaries~\cite{Ramirez:2022mys}, non-disruption of galaxy clusters~\cite{moore1993upper} and the non-observation of excess disk heating~\cite{Carr:2020gox} are shown in gray.}
\end{figure}
\subsection{Discussion}
In Figure~\ref{fig:FinalBounds} we collect our MACHO bounds and compare them to the bound derived in~\cite{Graham:2023unf} from Segue~I alone (purple) and a range of other MACHO bounds~\cite{EROS-2:2006ryy,Oguri:2017ock,Wilkinson:2001vv,10.1093/mnras/stac915,Ramirez:2022mys,moore1993upper,Carr:2020gox} (gray). We show the  bound from Ursa~Major~III in dot-dashed red to indicate that it cannot be considered a strict bound until Ursa~Major~III is confirmed as an UFD. In colored solid and dashed lines we also show the bounds for all UFDs in Figure~\ref{fig:UFDBounds}, assuming Dehnen and NFW DM profiles, respectively. The Figure clearly shows that the bound set by Segue~I which was already discussed in~\cite{Graham:2023unf} is both conservative and robust. On the one hand, there is a large number of confirmed UFDs which give comparable bounds. On the other hand bounds assuming a cuspy NFW DM halo are comparable albeit slightly stronger for lighter MACHO masses $\Mm \lesssim 10^4 M_\odot$. There is mounting evidence that compact UFDs and Segue~I in particular feature cuspy DM profiles~\cite{10.1093/mnras/sty2505,Hayashi_2023}, providing reason to believe that the NFW profile should be used to set bounds. Furthermore, should Ursa~Major~III be confirmed as an UFD, it has the potential to massively improve MACHO bounds in the $1M_\odot \leq \Mm \leq 10^5 M_\odot$ mass range and would constitute the strongest bound on MACHOs in this region of parameter space.  Also note that the MACHO bounds from Ursa~Major~III, reinterpreted for PBHs with masses $M_{\rm PBH} \lesssim 10^3 M_\odot$, are competitive and approximately equally strong as bounds on PBHs from gravitational waves~\cite{Kavanagh:2018ggo,LIGOScientific:2019kan,Chen:2019irf} and accretion~\cite{Serpico:2020ehh,Hektor:2018qqw,Manshanden:2018tze,Lu:2020bmd}. Even if Ursa~Major~III turns out to be a stellar cluster, there is an increasing number of ultra compact UFD candidates (see e.g.~\cite{Simon:2024lah}), with similar properties as Ursa~Major~III, which have the potential to improve bounds considerably.

\section{Conclusions}
In this work, we have extended an existing analysis~\cite{Graham:2023unf} of dynamical heating of stars in UFDs as a probe of dark matter in several ways: we quantified the impact of uncertainties in UFD parameters on heating based DM bounds and incorporated a wider set of UFD targets, including the recently discovered promising UFD candidate Ursa~Major~III.
This significantly enhances the robustness of existing bounds and highlights the unprecedented potential of Ursa~Major~III and similar recently-discovered ultra-compact and massive UFD candidates to significantly enhance the reach of stellar-heating based constraints.

We derived new limits on MACHO DM and studied how the choice of DM and stellar profile affects the bounds. We found that there is only a weak dependence on the stellar profiles that we considered (Plummer and Exponential profile) with the Plummer profile giving slightly more conservative bounds. The choice of DM profile (cuspy NFW or cored Dehnen) has opposite effects on different regions of parameter space. While cuspy halos enhance heating from lighter objects ($\Mm \lesssim 10^4 M_\odot$), cored halos are more sensitive to higher masses, implying that a cored Dehnen profile is not a conservative choice over the whole parameter space. We also assessed the impact of uncertainties in UFD parameters on the bound and found that due to its large uncertainty, varying the total halo mass $M_{\rm UFD}$ has the biggest effect on the bound, especially for $\Mm \gtrsim 10^5M_\odot$ where the bound is dominated by sampling limitations, i.e. UFD halos are not massive enough to contain at least $3$ MACHOs.  Thus, UFDs might not be the best systems to obtain reliable bounds for large MACHO masses. A promising complementary approach to cover and extend the large-mass region and partially circumvent the problem of large halo mass uncertainties would be to study stellar heating in globular clusters within massive galaxies, such as the Milky Way or the Magellanic Clouds~\cite{GCinProgress}. Nonetheless, the agreement of bounds across a large set of UFDs with the bound from Segue~I from~\cite{Graham:2023unf} provides confidence in the overall robustness of the bound.

We also presented new limits using the observation of Ursa~Major~III, which if confirmed as an UFD, would set the strongest limits to date on MACHOs and PBHs in the mass range $1M_\odot \leq \Mm \leq 10^5 M_\odot$. 

We presented, for the first time, projected limits based on the observation of Ursa~Major~III, which if confirmed as a UFD would yield the most stringent constraints to date on MACHOs and PBHs in the mass range $1M_\odot \leq \Mm \leq 10^5 M_\odot$. Thus, the sensitivity to substructure at previously inaccessible scales provides further motivation to pursue detailed observations of Ursa~Major~III specifically, and of the faintest UFDs more generally.

Our results underscore the power of UFDs as astrophysical laboratories for dark matter physics. As more UFDs are discovered and their kinematics measured with increasing precision, heating-based constraints will continue to improve, offering a robust probe of gravitationally interacting dark matter scenarios which remain inaccessible at terrestrial experiments.

\acknowledgments

We thank Ting Li, Ben Safdi and Oren Slone for useful discussions. This work was supported in part by NSF Grant No.~PHY-2310429, NSF Grant No.~PHY-2515007,  Simons Investigator Award No.~824870,  the Gordon and Betty Moore Foundation Grant No.~GBMF7946, the John Templeton Foundation Award No.~63595 and the University of Delaware Research Foundation.

\appendix
\section{MACHO and Stellar Scale Radius Dependence of Total Energy}\label{app:EnergyRadius}
In order to find how heating, i.e.~energy transfer, changes the scale radius of the distribution of a species $A$, one has to identify the dependence of the energy per unit mass of species $A$, i.e.~$\mathcal{E}_A$, on the scale radius. As was already discussed in Appendix~A of~\cite{Graham:2023unf}, identifying the energy of just one species is ambiguous since the gravitational potential energy also receives pairwise contributions, such that only the total potential energy is well-defined. Using the virial theorem ${\rm PE}_{\rm tot} = -2{\rm KE}_{\rm tot}$, which only holds for the sum over all species, the total energy takes the form
\begin{equation}\label{eq:Etot}
\begin{split}
    {\rm E}_{\rm tot} &= {\rm KE}_{\rm tot} + {\rm PE}_{\rm tot} =  -{\rm KE}_{\rm tot} \\
    &= -{\rm KE}_* - {\rm KE}_{\rm MACHO} - {\rm KE}_{\rm smooth}\,,
\end{split}
\end{equation}
where the kinetic energy of species $A$ is given by
\begin{equation}
    {\rm KE}_A = \frac{3}{2}\int_0^\infty 4\pi R^2\rho_A (R) \sigma_A^2(R)\, dR\,,
\end{equation}
where $\rho_A(R)$ is the density distribution of $A$ and $\sigma_A(R)$ is the velocity distribution for $A$ in hydrostatic equilibrium
\begin{equation}
    \sigma_A^2(R) = \frac{G}{\rho_A(R)}\int_R^\infty \rho_A (r) \frac{M_{\rm enc}(r)}{r^2}dr\,,
\end{equation}
with $M_{\rm enc}(r)$ being the mass enclosed in a sphere with radius $r$.
\begin{figure*}[t]
\centering
\includegraphics[width=0.49\textwidth]
{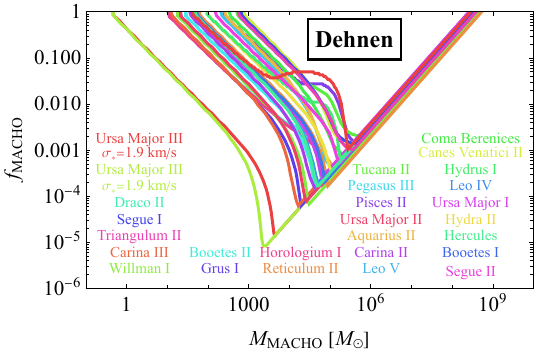}
\hfill
\includegraphics[width=0.49\textwidth]
{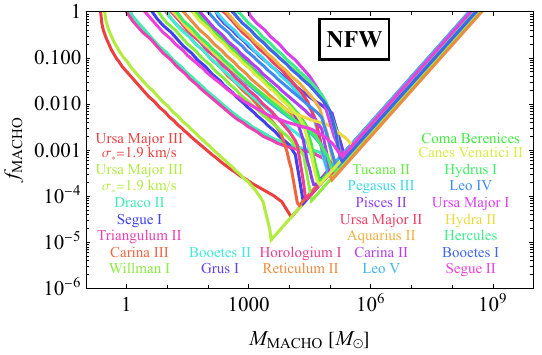}
\caption{\label{fig:AllUFD} Limits on the fraction of MACHO DM $f_{\rm MACHO}$ as a function of the MACHO mass $M_{\rm MACHO}$ from dynamical heating of stars for all UFDs shown in Table~\ref{tab:UFDList}. The {\bf left} panel shows the bounds assuming a Dehnen and the {\bf right} panel assuming an NFW DM profile. Both panels assume a Plummer sphere for the stellar density.}
\end{figure*}

It was argued in~\cite{Graham:2023unf} that under the assumption that the MACHO and smooth DM distributions remains constant as MACHOs heat the stars and $M_{\rm MACHO}^{\rm tot} =\fm M_{\rm UFD}\ll M_{\rm UFD}$, the $R_{0,*}$ dependence of the potential energy of the star-MACHO system is negligible. Thus, $\mathcal{E}_*$ can be defined as all remaining terms of Eq.~\eqref{eq:Etot} in the $\fm\rightarrow 0$ limit which are proportional to at least one power of $M_*$. Using this prescription and following the technique in~\cite{Graham:2023unf}, we have calculated $d\mathcal{E}_*/dt$ for all combinations of DM and stellar distributions used in this work.  To leading order in $1/R_{\rm UFD}$ these are given by
 \begin{widetext}
\begin{align}
&\text{Dehnen + Plummer:} &\frac{d\mathcal{E}_*}{dt} &= \frac{d R_{0, \star}}{d t} \frac{G M_*}{64} \left(\frac{3 \pi M_{\star}}{R_{0, \star}^2}+\frac{16\,M_{\mathrm{UFD}} R_{0, \star}}{R_{\mathrm{UFD}}^3}\left(6 \log \frac{R_{\mathrm{UFD}}}{R_{0, \star}}+(3 \log (4)-17)\right)\right)\,,\\
&\text{Dehnen + Exp:} &\frac{d\mathcal{E}_*}{dt} &= \frac{d R_{0,*}}{dt} \frac{G M_*}{64} \left( \frac{5 M_*}{R_{0,*}^2} + \frac{384 M_{\rm UFD} R_{0,*}}{R_{\rm UFD}^3} \right)\,,\\
&\text{NFW + Plummer:} &\frac{d\mathcal{E}_*}{dt} &= \frac{d R_{0,*}}{dt} \frac{G M_*}{64}\left( \frac{3\pi M_*}{R_{0,*}^2} + \frac{32 M_{\rm UFD} f(c)}{R_{\rm UFD}^2} \right)\,, \label{eq:EpsStarNFWPlum}\\
    &\text{NFW + Exp:} &\frac{d\mathcal{E}_*}{dt} &= \frac{d R_{0,*}}{dt} \frac{G M_*}{64}\left( \frac{5 M_*}{R_{0,*}^2} + \frac{48 M_{\rm UFD} f(c)}{R_{\rm UFD}^2} \right)\,,\label{eq:EpsStarNFWExp}
\end{align}
 \end{widetext}
where $f(c)^{-1} = \log(1+c) -c/(1+c)$.

For MACHO migration we need a similar expression for $\mathcal{E}_{\rm MACHO}$ as a function of $R_{\rm MACHO}$. Assuming that $R_{0,*}$ and $R_{\rm UFD}$ remain constant during migration, this is given by all terms in Eq.~\eqref{eq:Etot} which are proportional to at least one power of $M_{\rm MACHO}^{\rm tot}$. The general expression is given in Eq.~(A23) of~\cite{Graham:2023unf}. Using the explicit DM profiles it is straightforward to obtain $d\mathcal{E}_{\rm MACHO}/dt = (d\mathcal{E}_{\rm MACHO}/d R_{\rm MACHO}) dR_{\rm MACHO}/dt$.
%
%%%%%%%%%%%%%%%
\section{MACHO bounds for All UFDs}\label{app:AllUFDBound}
%%%%%%%%%%%%%%
%
Here we present the MACHO bounds of all UFDs in Table~\ref{tab:UFDList}, not just the strongest bounds. In Figure~\ref{fig:AllUFD} we show the bounds assuming a cored Dehnen and a cuspy NFW profile, in both cases assuming a Plummer stellar profile.
We see that Ursa Major III, if confirmed as an UFD, would give significantly stronger bounds than the other known UFDs.
\bibliography{biblio.bib}
\end{document}